\documentclass[useAMS,aasms,astrobib,usenatbib]{mn2e}
\usepackage{epsfig,amssymb,graphicx,rotating,color}
\usepackage{hyperref}
\usepackage{txfonts}

\title[A limit on eccentricity growth]{A limit on eccentricity growth from global 3-D simulations of disc-planet interactions}
\author[A. C. Dunhill, R. D. Alexander \& P. J. Armitage]{A. C. Dunhill$^{1}$\thanks{E-mail:
alex.dunhill@leicester.ac.uk}, R. D. Alexander$^{1}$ and P. J. Armitage$^{2,3}$\\
$^{1}$Department of Physics \& Astronomy, University of Leicester, Leicester, LE1 7RH\\
$^{2}$JILA, University of Colorado \& NIST, 440 UCB, Boulder, CO 80309-0440, USA\\
$^{3}$Department of Astrophysical and Planetary Sciences, University of Colorado, Boulder, USA\\}
\begin{document}
\voffset=-0.25in
\newcommand{\Msun}{$M_{\odot}$}
\newcommand{\Mjup}{$M_{\mathrm {Jup}}$}
\defcitealias{papaloizouetal01}{PNM01}

\date{Accepted 2012 October 19.  Received 2012 September 17; in original form 2012 June 6}

\pagerange{\pageref{firstpage}--\pageref{lastpage}} \pubyear{2012}

\maketitle

\label{firstpage}

\begin{abstract}
We present high resolution 3-D simulations of the planet-disc interaction using smoothed particle hydrodynamics, to investigate the possibility of driving eccentricity growth by this mechanism. For models with a given disc viscosity ($\alpha = 0.01$), we find that for small planet masses (a few Jupiter masses) and canonical surface densities, no significant eccentricity growth is seen over the duration of our simulations. This contrasts with the limiting case of large planet mass (over twenty Jupiter masses) and extremely high surface densities, where we find eccentricity growth in agreement with previously published results. We identify the cause of this as being a threshold surface density for a given planet mass below which eccentricity growth cannot be excited by this method. Further, the radial profile of the disc surface density is found to have a stronger effect on eccentricity growth than previously acknowledged, implying that care must be taken when contrasting results from different disc models. We discuss the implication of this result for real planets embedded in gaseous discs, and suggest that the disc-planet interaction does not contribute significantly to observed exoplanet eccentricities. 
\end{abstract}


\section{Introduction}\label{section1}

Modern theories of planet formation all posit that planets are the end result of the evolution of circumstellar discs of gas and dust, termed protoplanetary discs. This basic idea is consistent with both the observation of such discs around young, solar-like stars \citep*[e.g.,][]{sargentbeckwith87,haischetal01}, and the near co-planar, near circular orbits of the planets in the Solar System. However, observations of extra-solar planets made over the last 15 years have revealed that many planetary systems do not share the neat structure of our own. In particular, the abundance of planets with eccentric orbits was unexpected given this formation mechanism. The observed distribution of eccentricities is nearly uniform between eccentricity $e=0$ and $\sim0.6$, and stretches all the way up to $e \simeq 1$ \citep[e.g.][]{wrightetal11,kaneetal12}.

At small semimajor-axis ($a\lesssim0.1$au) there is a clear preference towards low eccentricity ($e<0.1$) orbits, and this is well explained by tidal circularisation \citep{rasioetal96}.  Similarly, the dearth of planets in high eccentricity orbits ($e\gtrsim0.5$) with small semi-major axes ($\lesssim1$au) is readily understood, as these planets have periastron distances $\sim 1\,R_{\odot}$ and consequently pass too close to their stars to be long-lived.  At larger semi-major axes, however, the picture is much less clear, and a number of scenarios have been proposed to to explain the apparent discrepancy between theory and observation. These fall into three broad categories: dynamical interactions of planets in multi-planet systems; secular interactions with companion stars; and tidal interactions between planets and their parent protoplanetary discs.

Large regions of the observed eccentricity distribution can be populated by invoking interactions between multiple planets, be it via direct close encounters \citep*{fordetal01,jurictremaine08,chatterjeeetal08} or through mean-motion resonances over longer time-scales \citep*{chiangetal02}. While simulations of such encounters are able to reproduce the observed distribution to a reasonable accuracy, it is unclear if such close interactions are frequent enough in nature to provide a universal source of planetary eccentricity. Another method for growing eccentricity is through secular interactions with inclined companion stars, which lead to a resonant exchange of angular momentum between the planets and the external body \citep{kozai62,lidov62}. This results in long-period changes in inclination and eccentricity and although this mechanism seems inviting as an alternative explanation for the observed eccentricity distribution, numerical work has shown that it does not produce the correct eccentricity distribtion \citep{takedarasio05}, although recent work has found that it may explain some eccentric misaligned Hot Jupiters \citep*{naozetal12}. Similar interactions between planets in the same system have also been suggested as a chaotic formation mechanism for highly eccentric planets \citep{wulithwick11}.

The gravitational interaction between a young embedded planet and its parent gas disc has also been suggested as a mechanism for driving eccentricity growth. For companions with masses comparable to the central body it has long been known that tidal interactions with the disc lead to eccentricity excitation.  This result has applications to stellar binaries \citep[e.g.,][]{artymowiczetal91} and binary super-massive black holes \citep[e.g,][]{cuadraetal09}, but how it extends to the more extreme mass ratios of star-planet systems is still not clear.

Analytical treatment of this problem begins with the consideration of resonant torques between a planet and its parent disc. These torques come in two main flavours, Lindblad and corotation. Lindblad torques occur where the epicyclic frequency in the disc is equal to plus or minus the frequency of a component of the planet's potential. Corotation resonances occur at radii where the component's orbital frequency matches that of the disc. For a planet with a circular orbit, the potential can be described by a series of components with the same angular velocity as the planet. This results in a coorbital corotation resonance with combs of Lindblad resonances to either side. When the planet's orbit is eccentric, this is no longer the case. A more complex picture emerges where not all resonances have the same angular velocity, resulting in non-coorbital corotation resonances \citep{goldreichtremaine80}. The interplay between these resonances, the planet that excites them, and how they are affected by various disc parameters have been discussed in detail by \citet{ogilvielubow03} and \citet{goldreichsari03}.  A great deal of nuance between competing effects is revealed, and in particular the possibility that corotation torques may begin to saturate and weaken once an initial eccentricity is attained is useful when considering the results of numerical work \citep{ogilvielubow03,massetogilvie04}.  However, the complex, non-linear nature of the planet-disc interaction makes the general problem analytically intractable, and numerical solution is required to gain a full understanding of eccentricity growth.

Semi-analytic calculations combining prescriptions from these analytic works have been somewhat inconclusive. \citet{moorheadadams08} found eccentric damping rather than growth in most cases, although in the cases where they did find growth it was extremely strong, leading to $e\sim1$ after only a few thousand orbits. However, as such highly eccentric planets would be unable to maintain an equally eccentric gap their orbits would be circularised as they interact with coorbital disc material \citep[e.g.][]{bitschkley10}.

By contrast, numerical simulations looking specifically at eccentricity growth have so far shown more consistent and positive results.  \citet*[][hereafter \citetalias{papaloizouetal01}]{papaloizouetal01} found that relatively massive embedded companions (in the brown dwarf regime) undergo eccentricity growth.  Lower-mass planets were not found to experience this growth, although it seems likely that this was for numerical rather than physical reasons \citep{massetogilvie04}. Later simulations have indeed found eccentricity growth, albeit at modest levels, down to \Mjup \citep*{dangeloetal06}.  Extensive analysis of the behaviour and morphology of the disc by \citetalias{papaloizouetal01} and \citet{kleydirksen06} attributes this eccentricity excitation to an instability launched at the 3:1 outer Lindblad resonance, which drives a large eccentricity at the inner edge of the disc. For the large companion masses considered by \citetalias{papaloizouetal01} a wide gap is opened in the disc, so coorbital co-rotation resonances are not present, and non-coorbital ones only operate once the planet's orbit is already eccentric. \citet{kleydirksen06} also extend this analysis down to planets of a few \Mjup, and find that this mechanism still operates down to planets of mass $\sim 3$ \Mjup, although the magnitude of the eccentricity induced depends strongly on the disc viscosity and temperature.

To date the majority of the numerical simulations of this problem have been performed in only two dimensions (2-D), and all have used Eulerian (grid-based) methods.  However, it has been suggested that a full three-dimensional (3-D) treatment weakens the effect of resonant torques \citep*{tanakaetal02}.  Each study has also typically only considered a single disc model, with little consistency in the choice of parameters, and it can be easily seen that the properties of the disc itself plays a large role in the evolution of a planet embedded within it.  Moreover, Eulerian methods are not always ideal for following the dynamics of gas on non-circular orbits; in general, one expects Lagrangian methods to track eccentric orbits with greater accuracy.

In this paper we present results of high-resolution 3-D smoothed particle hydrodynamics (SPH) simulations of eccentricity growth due to planet-disc interactions. In section \ref{section2} we describe our numerical method and the initial conditions used. In section \ref{section3} we present results of our simulations, comparing them to the results of \citetalias{papaloizouetal01} especially, and describing the effect of various model parameters. Section \ref{section4} contains a discussion of both the numerical limitations of our simulations, and the physical interpretation of the results. Our summary and concluding remarks are in section \ref{conclusion}.

\section{Simulations}\label{section2}

We have performed a suite of three-dimensional simulations of a planet embedded in a protoplanetary accretion disc. We used a modified version of the publicly available \textsc{Gadget-2} code \citep{springel05}, a hybrid SPH/N-body code. SPH methods are highly suited to tracing the evolution of dynamical systems, as the Lagrangian formulation means that important orbital parameters (energy, linear \& angular momentum) are naturally conserved \citep[e.g.,][]{price12}. We treat the star and planet as point masses, and model the gas disc using SPH. The basic parameters of the code are unchanged throughout our simulations, and were set as follows. The number of nearest neighbours for kernel integration is 50, and smoothing lengths are adjusted accordingly when the number of neighbours changes by more than $\pm 2$. The Courant parameter used to determine the time-step for SPH particles was set to $0.1$. The gravitational softening length for the point mass particles was in each case set to be the same as the sink radius for the planet particle (described below).  SPH particles are therefore accreted by the sink particles before they encounter gravitational softening, so this softening has no effect on our simulations.

We have altered the public \textsc{Gadget-2} code in a number of ways to make it more suitable for simulating planets embedded in discs. Firstly, we note that the standard Barnes-Hut tree is insufficiently accurate when computing the gravitational forces on the N-body particles (particularly the `planet'). Consequently, following the method of \citet{cuadraetal09}, we removed the N-body particles from the tree, and instead computed their gravitational forces by direct summation. As this is done for only two particles the additional computational cost is not large, and ensures high accuracy for the orbit of the planet. Additionally, following \citet{cuadraetal06}, we treat each N-body particle as a `sink' for SPH particles, accreting the mass and momentum of the swallowed particles on to the target sink, ensuring conservation of linear and angular momentum to a high accuracy. For the planet particle, the radius within which the sink will accrete SPH particles is set to be $0.4$ of the Hill radius, defined as
\begin{equation}
R_{\mathrm{Hill}} = a \left( \frac{M_p}{3M_{\star}}\right)^{1/3},
\label{eq1}
\end{equation}
for a planet and star of mass $M_p$ and $M_{\star}$ respectively with semimajor axis $a$. The sink radius for the star particle was set to be 7/8 of the initial inner disc radius.  We have also substituted the energy and entropy equations in \textsc{Gadget-2} with a locally isothermal equation of state, where the sound speed $c_{\mathrm s}$ in the disc is a function only of the cylindrical radius $R$.  We choose a power-law profile for the disc temperature $T(R) \propto R^{-1/2}$, consistent with both a linear viscosity law \citep[e.g.,][]{hartmannetal98} and a standard `flaring disc' model \citep[e.g.,][]{kenyonhartmann87}.

\subsection{Artificial viscosity and angular momentum transport}\label{section2.1}
In order to smooth out the discontinuities associated with shocks in SPH, it is necessary to make use of an artificial viscosity.  We use the method of \citet{morrismonaghan97} to dissipate energy in shocks and prevent particle penetration. This contributes a further acceleration term to the equations of motion \citep{springel05}\footnote{In this system of equations, $i$, $j$ and $k$ are vectors using summation notation, and $a$ and $b$ are particle indices.}
\begin{equation}
\frac{du^i_a}{dt} = -\sum_b m_b \Pi_{ab}\nabla^i_a \overline{W}_{ab}.
\label{eq2}
\end{equation}
$W_{ab}$ is the SPH kernel used by \textsc{Gadget-2}, a cubic spline that goes to zero at one smoothing length $h$, while $\overline{W}_{ab}$ is the arithmetic mean of  $W_{ab}(h_a)$ and $W_{ba}(h_b)$\,\footnote{Note that this differs from the `standard' SPH definition of the smoothing length, where the kernel goes to zero at $2h$.  This difference in notation has no practical effect, but must be considered when comparing these equations to those in the SPH literature.}. $\Pi_{ab}$ describes the strength of the artificial viscosity and is given by
\begin{equation}
\Pi_{ab} =\left \{ 
\begin{array}{ll}
(-\overline{\alpha}_{ab} \, \overline{c}_{s_{ab}}\, \mu_{ab} + \beta\, \mu_{ab}^2)/{\overline{\rho}_{ab}} & \mathrm{for\,\,}\bmath{u}_{ab} \cdot \bmath{r}_{ab} <0,\\
0& \mathrm{otherwise}.
\end{array}
\right.
\label{eq3}
\end{equation}
with $\mu$ given by
\begin{equation}
\mu_{ab}=\frac{\overline{h}_{ab}\, \bmath{u}_{ab} \cdot \bmath{r}_{ab}}{|\bmath{r}_{ab}|^2+\varepsilon\overline{h}_{ab}^{\,2}}.
\label{eq4}
\end{equation}
Here $\overline{c_s}_{ab}$, $\overline{\rho}_{ab}$ and $\overline{h}_{ab}$ represent the arithmetic means of the SPH sound speed, density and smoothing lengths respectively, between particle $a$ and $b$ while $\bmath{u}_{ab}$ and $\bmath{r}_{ab}$ are the differences between the velocity and position vectors of those particles.  $\varepsilon=0.0001$ is a numerical `safety factor', which prevents the artificial viscosity diverging for very small particle separations.

In the \citet{morrismonaghan97} scheme, $\beta$ in equation \ref{eq3} is given by $\beta = 2\overline{\alpha}_{ab}$, and $\overline{\alpha}_{ab}$ is the mean value $(\alpha_a + \alpha_b )/2$. We have used the \citet{price04} version of this method, where $\alpha_a$ is evolved according to
\begin{equation}
\dot{\alpha_a} = \frac{\alpha_{min} - \alpha_a}{\tau_a} +\left(\alpha_{max}-\alpha_a\right)S_a
\label{eq5}
\end{equation}
and $\alpha_{min}$ and $\alpha_{max}$ are imposed minima and maxima, set as a global parameters. The source term $S$ is given by
\begin{equation}
S_a=\mathrm{max} \{ -\bmath{\nabla} \cdot \bmath{u_a},\, 0\}
\label{eq6}
\end{equation}
and the decay time-scale is
\begin{equation}
\tau_a=\frac{h_a}{lc_{s_a}}.
\label{eq7}
\end{equation}
where $l$ is a dimensionless scaling parameter. We follow \citet{price04} and chose values $l=0.1$, $\alpha_{min}=0.01$ and $\alpha_{max}=2.0$.  We also make use of the `Balsara switch' \citep{balsara95} to limit the spurious detection of shear flow as shocks. 

This artificial viscosity prescription is necessary to handle shocks correctly, but we also wish to include an explicit physical viscosity to drive angular momentum transport in the disc.  To this effect we have modified the SPH equation of motion \citep[equation 7 in][]{springel05} to include a Navier-Stokes shear viscosity term \citep[with the bulk coefficient set to zero; ][]{lodatoprice10}:
\begin{equation}
\frac{du^i_a}{dt} = - \sum_b m_b \left[ \frac{S^{ij}_{a}}{\Lambda_a\rho_a^2} \nabla_a^j W_{ab}(h_a) + \frac{S^{ij}_{b}}{\Lambda_b\rho_b^2} \nabla_b^j W_{ab}(h_b)\right].
\label{eq8}
\end{equation}
Here $\Lambda_a$ is dimensionless and accounts for smoothing-length gradients:
\begin{equation}
\Lambda_a=\left(1+\frac{h_a}{3\rho_a} \frac{\upartial\rho_a}{\upartial h_a}\right)
\label{eq9}
\end{equation}
and $S^{ij}_a$ is the stress tensor
\begin{equation}
S^{ij}_a=\left[-(P_a) + \left(\frac{2}{3}\eta_a\right) \frac{\upartial u_a^k}{\upartial r_a^k}\right]\delta^{ij} + \eta \left(\frac{\upartial u_a^i}{\upartial r_a^j} + \frac{\upartial u_a^j}{\upartial r_a^i}\right).
\label{eq10}
\end{equation}
$P$ and $\rho$ are the SPH-estimated pressure and density respectively. The standard variable-smoothing-length operator is given by \citep{lodatoprice10}
\begin{equation}
\frac{\upartial u_a^i}{\upartial r_a^j}=\frac{1}{\Lambda_a\rho_a}\sum_b m_b \left(u^i_a-u^i_b \right) \frac{\upartial W_{ab}(h_a)}{\upartial r^j_a}.
\label{eq11}
\end{equation}
$\eta_a$ is the shear viscosity, which is parametrized in terms of kinematic shear viscosity $\nu_a$ by 
\begin{equation}
\nu_a = \eta_a/\rho_a.
\label{eq12}
\end{equation}
In practice we set the viscosity such that each particle $a$ has a viscosity $\nu_a$ which depends on only its (cylindrical) radius, which gives a straightforward means of implementing the power-law viscosity prescriptions described in section \ref{section2.2}.

\subsubsection{Viscous ring-spreading tests}\label{section2.1.1}

\begin{figure*}
\includegraphics[width=\linewidth]{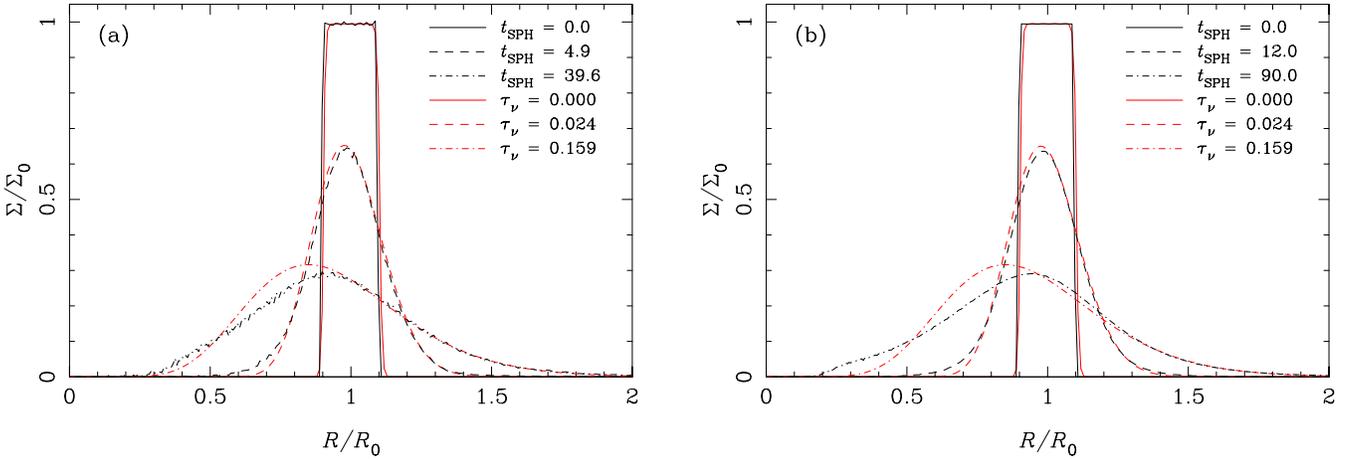}
\caption{Radial surface density evolution of a viscously spreading ring. Panel (a) shows a ring with $10^5$ particles in its initial configuration (solid black line), after 4.9 orbits (dashed black line) and after 39.6 orbits (dash-dotted black line). The corresponding red lines show the best-fitting profile from an explicit 1-D ring code, plotted as a fraction of the viscous spreading time. Panel (b) is as (a), but for a ring with $10^6$ particles, in its initial configuration (solid line), after 12 orbits (dashed line) and after 90 orbits (dash-dotted line).  Our Navier-Stokes viscosity approximates a uniform viscosity well except at very low resolution, where the `best fit' to the 1-D ring is rather poor in both cases.  The initial configuration in panel (b) is representative of the resolution we obtain in our full disc models described in section \ref{section2.2}.}
\label{fig1}
\end{figure*}

To test this set-up, we conducted test simulations which model the viscous spreading of a gas ring around a point mass.  In these tests a thin ring of initially uniform surface density is allowed to evolve under the action of viscous torques.  We expect the ring to spread \citep[see, e.g.,][]{pringle81}, and by comparing our results against those from a one-dimensional explicit scheme \citep*[e.g.,][]{pringleetal86}  we can test the accuracy of our SPH viscosity prescription. We modelled the ring using $10^5$ and $10^6$ SPH particles and four different levels of viscosity (see table \ref{table1}). The ring was set up orbiting a single point mass, and was allowed to evolve for 200 orbits.

The viscous time-scale for a spreading ring with constant viscosity is 
is given by \citep{pringle81}
\begin{equation}
\tau_{\nu} = \frac{R_0^2}{12\nu} \, .
\label{eq13}
\end{equation}
We measure the effective viscosity in the SPH simulations by fitting the resulting surface density profiles $\Sigma(R,t)$ to those obtained from the 1-D explicit code.  We perform a least-squares fit for $\Sigma(R)$ at several different times in each simulation, and measure the effective viscous time-scale $\tau_{\nu}$. The ring spreading is initially linear with the 1-D model as expected, but later the approximation of constant viscosity becomes invalid as the artificial viscosity becomes strong at the ring edges. We only fit spreading times during this initial phase where the linear relationship exists. We are thus able to compare the imposed viscosity $\nu_{\mathrm {in}}$ with the measured rate of viscous angular momentum transport $\nu_{\mathrm {out}}$, in order to determine the accuracy of our numerical scheme.  We further parametrize the viscosity in terms of an effective \citet{shakurasunyaev73} $\alpha$ parameter by assuming that  $\nu=\alpha c_{s0} H_{0}$ at $R_0$.  Our measured values are given in table \ref{table1}, and typical best fits are shown in figure \ref{fig1} (for the runs with $\nu_{\mathrm{in}}=10^{-5}$ at different fractions of the viscous timescale $\tau_{\nu}$).  

From these results we can estimate the true level of angular momentum transport present in our disc simulations. We see from Table \ref{table1} that for both the 
$10^5$- and $10^6$-particle runs, there is essentially no difference in the measured viscosity between the runs with $\nu_{\mathrm {in}} = 0$ (i.e., artificial viscosity only) and $\nu_{\mathrm {in}} = 10^{-5}$, and that the viscosity in the higher-resolution runs is smaller than that in the lower-resolution runs (by a factor of approximately $10^{1/3} \simeq 2.15$), suggesting that in this regime artificial transport of angular momentum is dominant.  For larger values of $\nu_{\mathrm {in}}$, however, the angular momentum transport increases as expected, showing that our imposed Navier-Stokes viscosity is the dominant source of angular momentum transport for $\nu_{\mathrm {in}} \gtrsim 10^{-5}$ (or, equivalently, $\alpha \gtrsim 0.008$).  An input $\alpha$ of 0.01 gives a value of $\nu$ at $R_0$ of $1.5 \times 10^{-4}$. The SPH smoothing lengths throughout our simulated discs are comparable to those in our $10^6$-particle spreading rings at radius $R_0$, being of order 0.01 in code units, and so we use this set of rings for comparison. This suggests that the artificial viscosity sets a floor to the effective viscosity in the SPH simulations, approximately at or slightly below our canonical imposed value of $\alpha = 0.01$. We are therefore satisfied that artificial transport of angular momentum does not dominate the viscosity in our disc models\footnote{The exception is run PNM, which uses a much lower explicit viscosity.  In this case we expect the artificial viscosity to dominate the angular momentum transport.}.  Moreover, it is known that  in the case of a shearing disc, the SPH artificial viscosity behaves similarly to a Shakura-Sunyaev $\alpha$ viscosity \citep{murray96}. Consequently, although the SPH artificial viscosity prevents us from running simulations with very low disc viscosities, we are confident that numerical angular momentum transport does not dominate our results.

\begin{table}
\begin{minipage}[t]{\columnwidth}\centering
\caption{Summary of ring spreading tests.  $\nu_{\mathrm {in}}$ denotes the magnitude of the imposed Navier-Stokes viscosity, as specified in equations \ref{eq10} and \ref{eq12}, while $\nu_{\mathrm {out}}$ is the measured viscosity in the SPH runs, calculated by fitting the viscous time-scale as in Equation \ref{eq13}.  The effective $\alpha$ values are calculated by assuming that $\nu_{out}=\alpha c_{s0} H_{0}$.  For very small values of $\nu_{\mathrm {in}}$ the artificial viscosity is the dominant source of angular momentum transport, but for $\nu_{\mathrm {in}} \gtrsim 10^{-5}$ the measured viscosity increases as expected.}\label{table1}
\begin{tabular}{cccc}
\hline
$N$ & $\nu_{\mathrm{in}}$ & $\nu_{\mathrm{out}}$ & Corresponding $\alpha$ \\
\hline
$10^5$ & $0$ & $3.01 \times 10^{-4}$ & $0.019$\\
$10^5$ & $10^{-5}$ & $3.06 \times 10^{-4}$ & $0.020$\\
$10^5$ & $10^{-4}$ & $3.45 \times 10^{-4}$ & $0.022$\\
$10^5$ & $10^{-3}$ & $6.92 \times 10^{-4}$ & $0.045$\\
$10^6$ & $0$ & $1.29 \times 10^{-4}$ & $0.008$\\
$10^6$ & $10^{-5}$ & $1.38 \times 10^{-4}$ & $0.009$\\
$10^6$ & $10^{-4}$ & $2.25\times 10^{-4}$ & 0.015\\
$10^6$ & $10^{-3}$ & $1.05 \times 10^{-3}$ & $0.068$\\
\hline
\end{tabular}
\end{minipage}
\end{table}

\subsection{Numerical set-up and initial conditions}\label{section2.2}
We choose a system of units (mass $M_0$, distance $R_0$ and time $P_0$) such that for a planet mass $M_P$ and a stellar mass $M_{\star}$, $M_P+ M_{\star} = 1 M_0$.  The unit of time $P_0$ is the Keplerian orbital period for a semi-major axis $R_0$. This choice of units fixes the gravitational constant to be $G=4\pi^2$. This system is particularly convenient as for a mass $M_0=1 M_{\odot}$ and radius $R_0=1$au it gives $P_0=1$ year. 

Our initial conditions consist of a gas disc which is axisymmetric about the centre of mass and extends radially from 0.4 to 6 $R_0$. It has a power-law surface density such that 
\begin{equation}
\Sigma(R) = \Sigma_0\left( \frac{R}{R_0}\right)^{-\gamma}.
\label{eq14}
\end{equation}
where $\Sigma_0 = \Sigma(R_0)$ is a reference surface density used for normalisation.  The viscosity $\nu$ in equation \ref{eq12} for each disc model (barring that used in run {\sc PNM}, see table \ref{table2}) was chosen such that $\nu\Sigma$ was constant (i.e., the disc is a steady-state accretion disc), and normalised such that ${\nu}_{0} = 0.01 {c_s}_0 H_0$ (where the subscript $0$ indicates values at $R_0$). This treatment reduces to a \citet{shakurasunyaev73} alpha-prescription, with $\alpha = 0.01$, in the canonical case of a $\Sigma \propto R^{-1}$ surface density profile.  The vertical scale-height $H$ is determined by the $T \propto R^{-1/2}$ temperature profile described above, so our disc has $H/R \propto R^{5/4}$. We normalise the temperature profile by setting $H/R=0.05$ at $R=R_0$.

\begin{table}
\begin{minipage}[t]{\columnwidth}\centering
\caption{Summary of parameters used in our disc models.}\label{table2}
\begin{tabular}{ccccl}
\hline
$\Sigma_0$ [Code Units]\,\footnote{As the unit of mass depends on the planet mass $M_P$, different values of $q$ give a different $\Sigma_0$ for the same physical model.} & $\Sigma_0\,[\mathrm{g/cm^2}]$\,\footnote{These values correspond to a 1\Msun\ star and an initial semi-major axis of $1$ au.} & $\gamma$ & $q$ & Model Name\\
\hline
$1.12 \times 10^{-5}$ & $10^2$ & $1$ & $0.005$ & {\sc Low5}\\
$1.10 \times 10^{-5}$ & $10^2$ & $1$ & $0.025$ & {\sc Low25}\\
$1.12 \times 10^{-4}$ & $10^3$ & $1$ & $0.005$ & {\sc High5}\\
$1.10 \times 10^{-4}$ & $10^3$ & $1$ & $0.025$ & {\sc High25}\\
$1.10 \times 10^{-4}$ & $10^3$  & $0$ & $0.025$ & {\sc Flat}\\
$7.03 \times 10^{-4}$ & $6.4 \times10^3$ & $0$ & $0.025$ & {\sc PNM}\,\footnote{Corresponds to the disc model used in \citetalias{papaloizouetal01}. For this model the viscosity $\nu$ in equation \ref{eq12} was $1.59 \times 10^{-6}$ in dimensionless units, far less than the artificial viscosity, see sections \ref{section2.1.1} \& \ref{section3.1.2}. This was set in error, and should have been $2.68 \times 10^{-5}$ in our units to match that used in \citetalias{papaloizouetal01}.}\\
$7.03 \times 10^{-4}$ & $6.4 \times10^3$ & $1$ & $0.025$ & {\sc PNMslope}\\
\hline
\end{tabular}
\end{minipage}
\end{table}

In each case our disc is modelled using $N=10^7$ SPH particles. We build the initial conditions by randomly distributing the particles in the radial and azimuthal directions according to equation \ref{eq14}. Particles were given zero velocity in the radial and vertical ($z$) directions, and azimuthal velocities $v_{\phi}$ such that
\begin{equation}
\frac{v_{\phi}^2}{R}= \frac{GM_{\star}}{R^2}+\frac{1}{\rho}\frac{\upartial P}{\upartial R} \, .
\label{eq15}
\end{equation}
(i.e., Keplerian orbital velocities, with a small correction to account for radial pressure gradients). The particles were distributed in the vertical direction by randomly sampling a Gaussian density profile with scale-height $H$. This set-up is not strictly in equilibrium, due to the discontinuities at the inner and outer disc edges, and consequently transient density waves pass through the disc for approximately one outer orbital period ($\sim15P_0$). These transients are short-lived, however, and tests performed with an initially relaxed disc show that these do not affect our results in any significant way.

We ran 7 models, using the disc parameters described in table \ref{table2} and using planet masses $q = M_P/M_{\star}$ = 0.005 \& 0.025. The star and planet were initially placed on the $x$-axis and given Keplerian orbits around the centre of mass, and each model was allowed to evolve for 200 planetary orbital periods (unless otherwise stated). The simulations were run on the {\sc alice} HPC cluster at the University of Leicester\footnote{See \url{http://go.le.ac.uk/alice}}, using 128 parallel cores.

\section{Results}\label{section3}

\subsection{Code tests}\label{section3.1}

\subsubsection{Numerical eccentricity damping}\label{section3.1.1}
\begin{figure}
\includegraphics[width=\columnwidth]{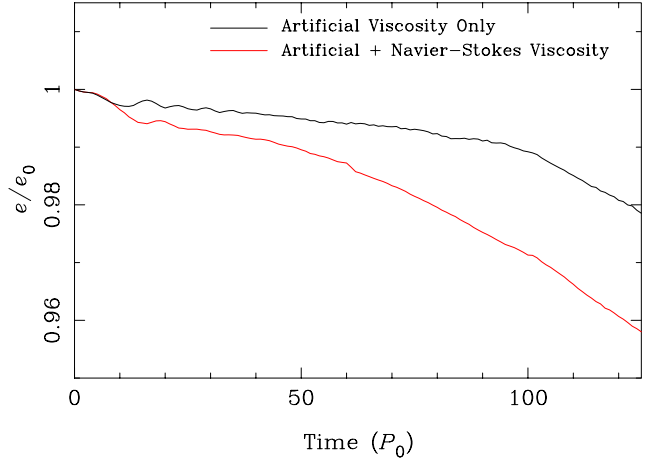}
\caption{Comparison between initially eccentric models with just artificial viscosity (black line) and with both artificial viscosity and a Navier-Stokes viscosity (red line). Both used disc {\sc Low25} (see table \ref{table2}) and were given an initial eccentricity $e_0=0.05$. The model without the Navier-Stokes viscosity shows small initial damping of eccentricity which soon flattens off, while the model with the full viscosity scheme implemented sees continued and increased eccentricity damping as it evolves. This shows that the SPH artificial viscosity is not causing spurious eccentricity damping. In the case of the full Navier-Stokes viscosity model, at later times the eccentricity decay reversed and began to grow again. This is due to the eccentric planet causing stronger disc eccentricity, and is in agreement with the findings of \citet{dangeloetal06}.}
\label{fig2}
\end{figure}

As an initial test, and to ensure that the effects we see are physical, we first verified that a planet on an eccentric orbit does not undergo spurious eccentricity damping due to the SPH artificial viscosity (or other numerical effects). To this end we ran 2 realisations the disc model {\sc Low25}, where in each case the planet was given an initial eccentricity $e_0=0.05$. In one version of this model the Navier-Stokes prescription described in equations \ref{eq8} to \ref{eq10} was switched off, so that in this case the only source of angular momentum transport was from the artificial viscosity. These were allowed to evolve for $125$ orbits, and the eccentricity evolution is shown in figure \ref{fig2}.

\begin{figure*}
\includegraphics[width=\linewidth]{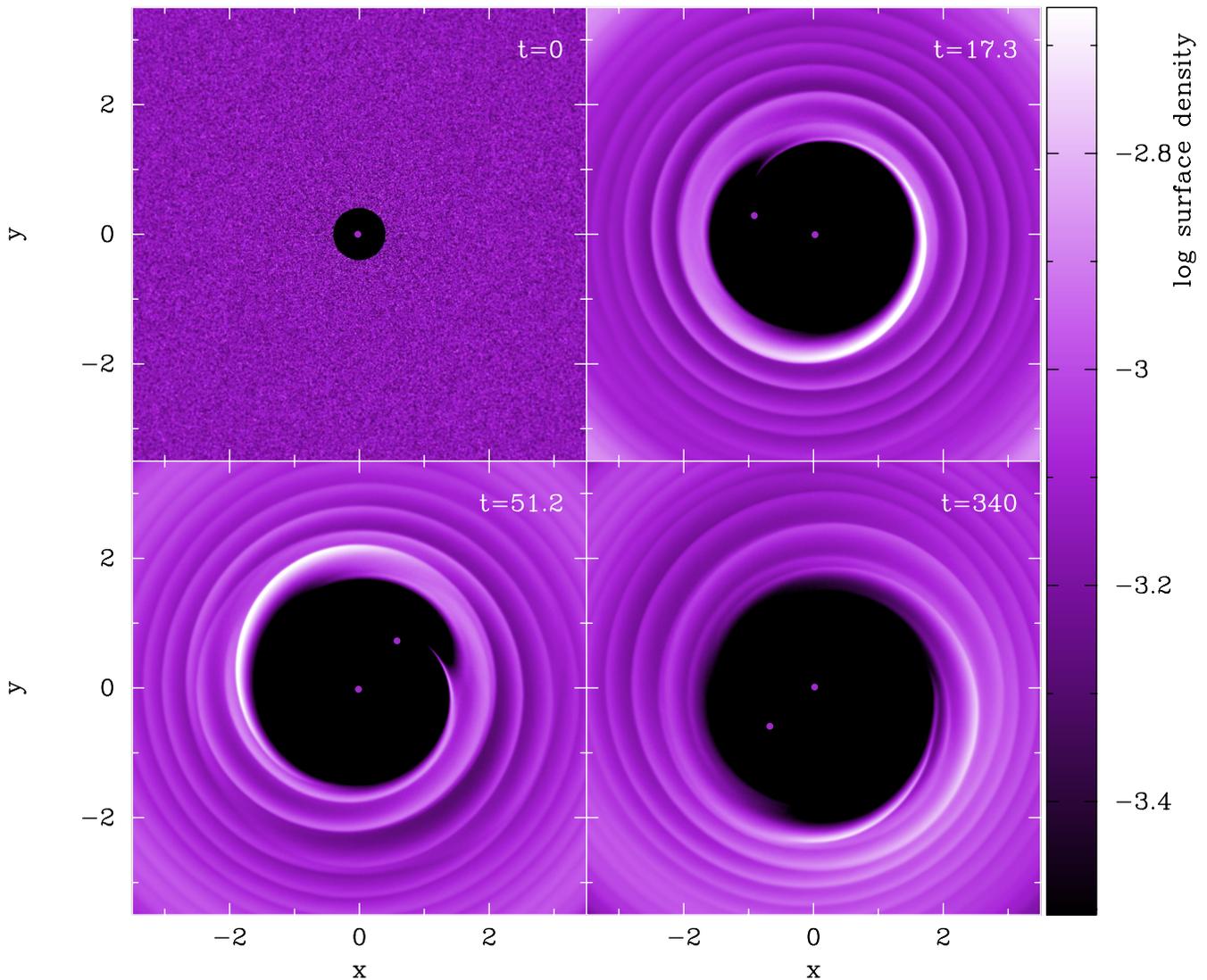}
\caption{Surface density evolution of the central region of our disc model {\sc PNM} (see table \ref{table2}), roughly equivalent to run N4 from \citetalias{papaloizouetal01}. The distance unit is equal to the initial separation between the star and planet. Times shown are in units of the initial orbital period of the planet. The eccentricity evolution for this model is shown in figure \ref{fig4}. After the inner disc clears (upper panels), the presence of the planet drives the inner edge of the disc eccentric (bottom panels). As the disc evolves it exerts torques back upon the planet, causing the planet's eccentricity to grow.  [All of the SPH maps presented here were rendered using \textsc{splash} \citep{price07}.]}
\label{fig3}
\end{figure*}

In the model with no physical (Navier-Stokes) viscosity we see very little initial eccentricity decay during the initial period while the disc settles into an equilibrium state and the planet opens its full gap. A loss of some eccentricity during this phase is fully expected and is in agreement with simulations by \citet{bitschkley10}, who found that for non-gap opening planets eccentricity is damped by the surrounding gas. With the physical viscosity switched on we see additional damping of eccentricity, at a much more pronounced level. In both cases we see exponential damping after the initial phase (beyond $\sim 75$ orbits), after the planet has fully cleared its gap. The rate of eccentricity damping during this phase is similar between the two models. This is because once the gap has fully formed, the planet is not directly interacting with the gas to any great extent so the angular momentum exchange here is due to gravitational resonances. At later times, the eccentricity began to rise again in the case of the full viscosity model. This is expected, as \citet{dangeloetal06} found that even for very low planet masses an initially eccentric planet can undergo far stronger eccentricity growth than one on an initially circular orbit. Consequently we conclude that the SPH artificial viscosity is not causing significant spurious eccentricity damping in our disc models.

\subsubsection{\citetalias{papaloizouetal01} Result}\label{section3.1.2}

\begin{figure}
\includegraphics[width=\columnwidth]{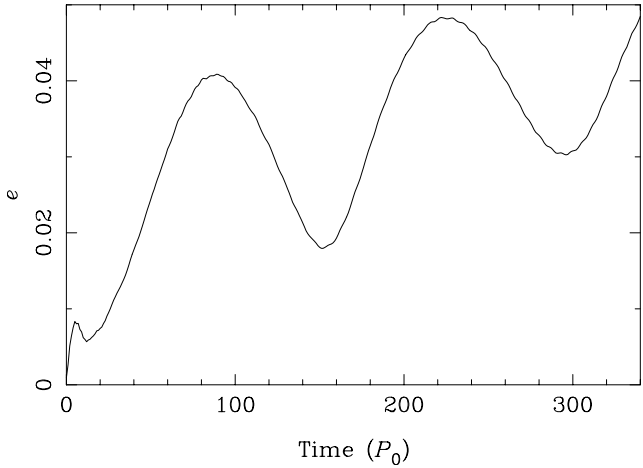}
\caption{Evolution of  planet eccentricity for the disc model {\sc PNM}, which is approximately equivalent to run N4 from \citetalias{papaloizouetal01} (see table \ref{table2}). We find growth of eccentricity in general agreement with that paper. Surface density plots from this run are shown in figure \ref{fig3}. The $\sim 100$ orbital period oscillations are due to the relative precession of the planet and the eccentric inner edge of the disc.}
\label{fig4}
\end{figure}

As a further test, we have also attempted to reproduce the results of  \citetalias{papaloizouetal01}. To this end we have created a disc model that is as near as possible in form to that used in their calculations\footnote{\citetalias{papaloizouetal01} used a 2-D fixed-grid code for their simulations, so it is not possible for us to run a completely identical simulation.}. Using the parameters for run {\sc PNM} given in table \ref{table2} this approximates run N4 from that paper, with the obvious caveat that our simulations are in 3-D. Unlike our other models, the normalisation for the viscosity law described in section \ref{section2.2} was taken to be $1.59 \times 10^{-6}$ in our dimensionless code units\footnote{Equivalent to $2.50 \times 10^{-6}$ in the units of \citetalias{papaloizouetal01}.}. This is constant across the disc, fulfilling the steady-state accretion requirement that $\nu \Sigma$ be constant. Note, however, that with this setup the angular momentum transport due to the explicit viscosity is smaller than that due to the SPH artificial viscosity (see Section \ref{section2.1.1}), so in practice our test calculation is somewhat more viscous than that of PNM01 (by a factor of a 2--3).  This model was allowed to evolve for 340 orbital periods. The evolution of the planet's eccentricity is shown in figure \ref{fig4}, while a series of surface density maps as the disc evolves are shown in figure \ref{fig3}. 

The evolution of our disc structure is broadly in line with that found by \citetalias{papaloizouetal01}, with the planet rapidly opening a wide gap in the disc, and the inner part of the disc quickly accreting onto the central star. As the system evolves the planet begins to drive eccentricity in the disc at its inner edge, while its own orbit remains essentially circular. At later times this is no longer the case and the planet's orbit becomes significantly eccentric [above the $\sim 0.01$-$0.05$ level required by \citet{ogilvielubow03} for non-coorbital corotation resonances to saturate, at which point further eccentricity growth is expected].  The long-period oscillations in eccentricity seen in figure \ref{fig4} are due to the relative precession of the planet and the eccentric disc inner edge of the disc.

The level of eccentricity growth seen in our simulations is somewhat less than that seen by \citetalias{papaloizouetal01}, but it is broadly comparable.  Moreover, given our larger effective disc viscosity, and the inherent differences between the methods (2-D fixed-grid versus our full 3-D SPH calculation), we do not expect exact agreement.  We also note that our simulations are extremely computationally expensive (using up to approximately 150,000 CPU hours per run), so we are limited in how long we can evolve our models for. Consequently we are unable to find a level of eccentricity at which growth saturates (the eccentricity was still growing at the end of our simulation), but otherwise we find good agreement between our 3-D results and the 2-D simulations of \citetalias{papaloizouetal01}.

\subsection{The effect of surface density \& planet mass}\label{section3.2}

\subsubsection{Planet mass}\label{section3.2.1}

\begin{figure}
\includegraphics[width=\columnwidth]{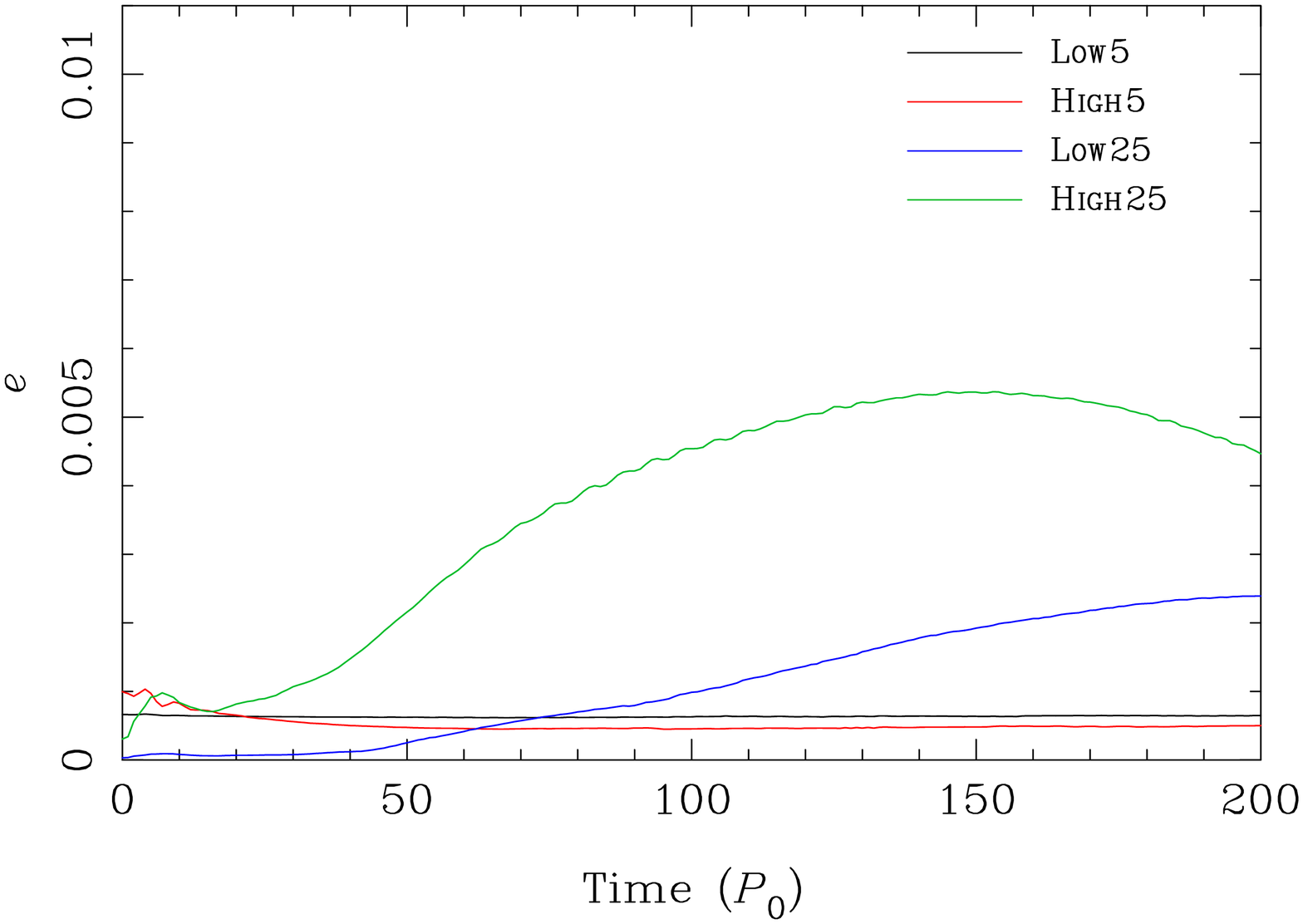}
\caption{Eccentricity evolution for planets of different masses in disc models with different surface densities. {\sc Low5} and {\sc High5} have a planet-star mass ratio of $q = 0.005$, while {\sc Low25} and {\sc High25} have $q=0.025$. {\sc Low} and {\sc High} refer to the choice of disc surface density -- see table \ref{table2} for the values -- but all have a power-law index of $\gamma=1$.}
\label{fig5}
\end{figure}

To test the effect of different planet masses, we ran models with both {\sc Low} and {\sc High} surface densities (see table \ref{table2}) with planet-star mass ratios of $q=0.005$ \& $0.025$. For a 1\Msun\ star this corresponds to planet masses of approximately 5 and 25\Mjup\ respectively. Figure \ref{fig5} shows the time evolution of the orbital eccentricity in each case. For both of the models with $q=0.005$ ({\sc Low5} and {\sc High5}), no eccentricity growth was seen at any level. For the models with $q=0.025$ ({\sc Low25} and {\sc High25}) some modest growth was seen, but none above $e=0.005$.

\begin{figure}
\includegraphics[width=\columnwidth]{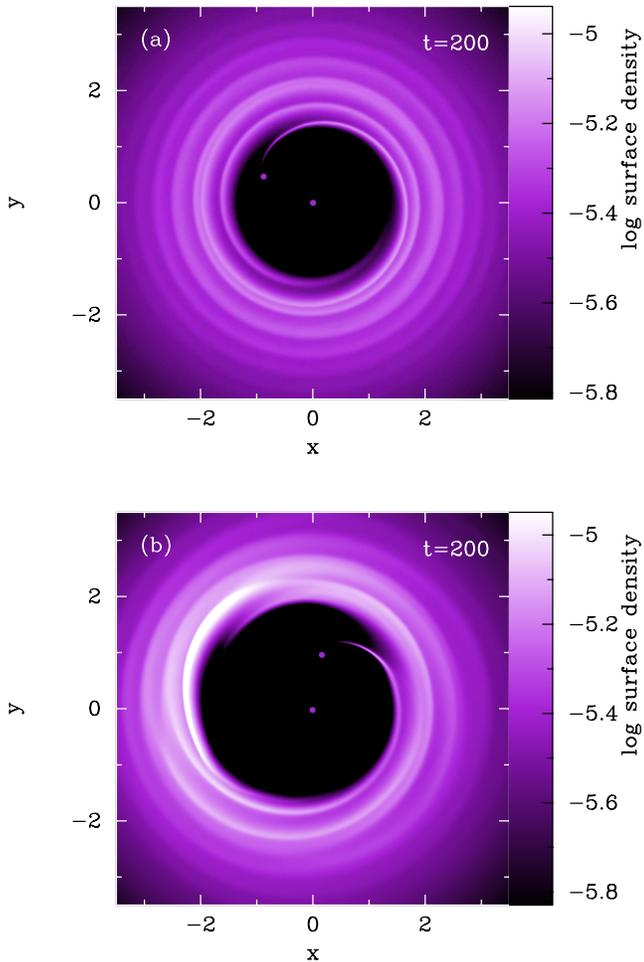}
\caption{Surface density maps after 200 planetary orbits: panel (a) shows the central region of model {\sc Low5} ($q = 0.005$) and panel (b) shows {\sc Low25} ($q = 0.025$; see table \ref{table2}). It is clear from panel (a) that the lower mass planet has a far smaller perturbing effect on its host disc, driving spiral waves but not further disrupting the disc shape or structure. This is in stark contrast to panel (b), where the higher mass planet has driven the inner edge of the disc quite eccentric.}
\label{fig6}
\end{figure}

This is consistent with the conclusions of both \citetalias{papaloizouetal01} and \citet{dangeloetal06}, who found that eccentricity is first induced in the disc, driven by the outer 3:1 Lindblad resonance. We clearly see this disc eccentricity in both runs with $q=0.025$, but not in either of the lower mass planet cases. A comparison of the surface density maps for both planet masses in the {\sc Low} model is shown in figure \ref{fig6}.  The reason for the lack of eccentricity growth in both of the $q=0.005$ models is that the 3:1 outer Lindblad resonance induced by the planet is too weak to affect the disc structure significantly. The difference in eccentricity evolution between models {\sc Low25} and {\sc High25} arises because the higher surface density disc is simply more massive, and consequently is able to exert stronger torques upon the planet, resulting in more (but still extremely limited) eccentricity growth.

We are confident that in the case of both planet masses, eccentric co-rotation resonances are resolved where present. Using the width formula provided by \citet{massetogilvie04}, our smoothing lengths are smaller than the resonant widths at the nominal resonant locations by a factor of at least two, for resonances that are not fully in the open gap. For the case of the 5 \Mjup planet, this is assuming an eccentricity of $10^{-2}$, as for the eccentricity measured from the simulations the prescribed widths are vanishing.

\subsubsection{Radial surface density profile}\label{section3.2.2}

\begin{figure}
\includegraphics[width=\columnwidth]{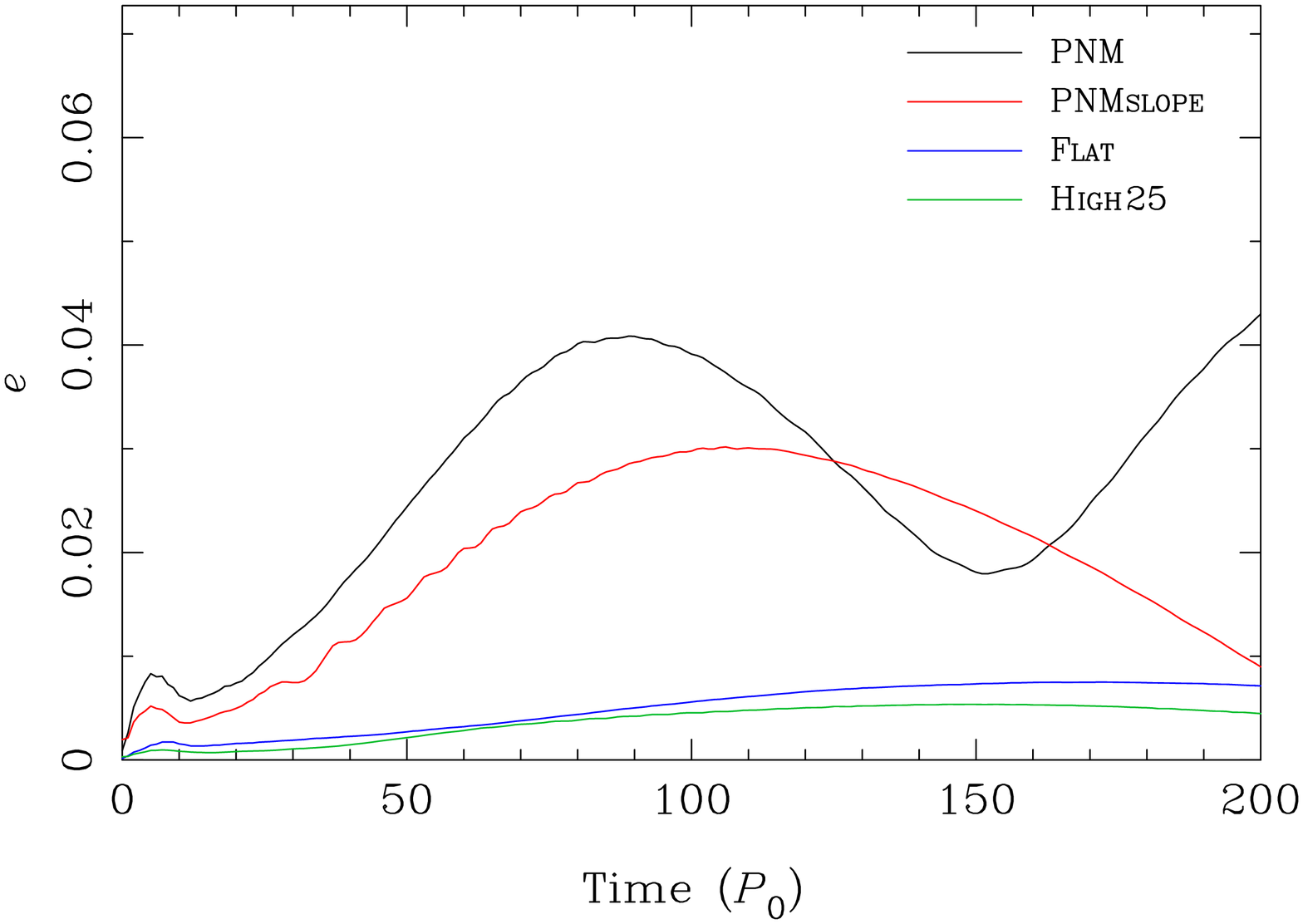}
\caption{Eccentricity evolution for disc models with different surface density profiles. The black line is the same as in figure \ref{fig4}, truncated at $t=200P_0$ for reference, and the green line is the same as in figure \ref{fig5}. These models show that while higher values of $\Sigma_0$ give consistently stronger eccentricity growth, shallower radial profiles (lower values of $\gamma$, see table \ref{table2}) also show stronger growth.  The oscillations in the curves are again due to the relative precession of the planet and the eccentric disc.  Note also that we ran model {\sc PNMslope} (red line) for a further 100 orbits (not shown here) to verify that the eccentricity does grow indeed continue to grow; the declining eccentricity at $t=200$ is simply due to this precession effect.}
\label{fig7}
\end{figure}

We now consider the effect of the disc surface density profile on the evolution of the embedded planet. To this end we have run two additional models with the same star-planet mass ratio of $q=0.025$, and different radial surface density profiles ($\gamma = 1$ \& 0). The disc models were {\sc Flat} and {\sc PNMslope} (see table \ref{table2}), and each was evolved for 200 orbits. The evolution of the orbital eccentricity for these runs, along with two previously described ({\sc High25} and {\sc PNM}, for the purposes of comparison), are shown in figure \ref{fig7}.  Again, due to computational limitations we have not run these models for long enough to determine at what level of eccentricity growth saturates; we are instead more concerned here with determining the conditions under which growth will occur. We see again the eccentric inner disc driven by the presence of the massive planet, as described above, and again find that its effect on the eccentricity of the planet depends strongly on the precise disc model used.

\begin{figure}
\includegraphics[width=\columnwidth]{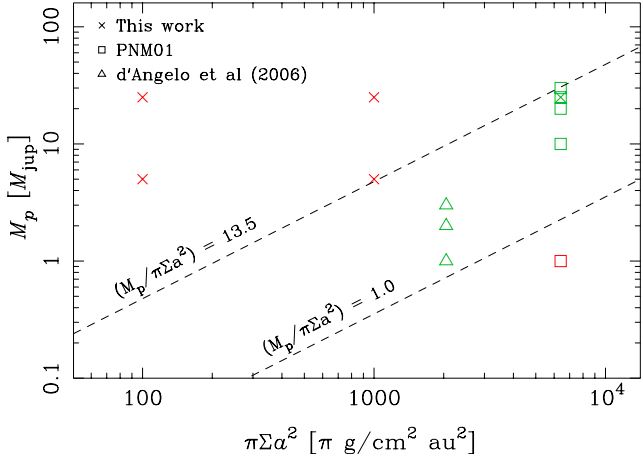}
\caption{Comparison between the local unperturbed disc mass (approximated by $\pi \Sigma a^2$) and planet mass for simulations presented here and in two other papers. Crosses indicate simulations from this paper, squares from \citetalias{papaloizouetal01} and triangles from \citet{dangeloetal06}. Green symbols indicate that eccentricity growth was seen, while red indicates that it was not. The choice of units on the x-axis gives the reference surface density at the semimajor-axis of the planet for that model (except for the models of \citet{dangeloetal06}, which placed their planets at 5.2au; in this case the x-axis value is $5.2^2$ times their reference surface density). The dashed lines show where the ratio of $M_p / \pi \Sigma a^2$ cross values of 1.0 and 13.5. We suggest that these values represent cases where the planet is massive enough to significantly perturb the inner edge of the disc, but where the disc is also massive enough that it can exert sufficiently strong torques upon the planet, although we note that this ratio is not the lone deciding factor. We also draw attention to the fact that the single point below the threshold $M_p / \pi \Sigma a^2 =1.0$ has been questioned by \citet{massetogilvie04} as being affected by spurious numerical factors, and thus the lower limit may not be real. We do not expect this to extend into the sub-jovian regime, where gap-opening is not necessarily efficient enough to allow similar modes of eccentricity growth (although this also depends on other factors including disc viscosity).}
\label{fig8}
\end{figure}

The results from these models show two things. First, the magnitude of the disc surface density has a strong effect on the eccentricity evolution of an embedded planet. Figure \ref{fig7} clearly shows that the two models with $\Sigma_0=7.03\times10^{-4}$ ({\sc PNM} and {\sc PNMslope}) see significant eccentricity growth, while the models with $\Sigma_0=1.10\times10^{-4}$ ({\sc High25} and {\sc Flat}) do not. This suggests that eccentricity can only be excited above a threshold surface density. This behaviour was suggested by \citetalias{papaloizouetal01} but not investigated in detail. Comparing the ratio between the planet mass and the local disc mass (approximated by $\Sigma \pi a^2$ in the unperturbed disc) suggests that values between $\sim1-13.5$ may result in eccentricity growth (see figure \ref{fig8}). We also note in passing that this is a surprisingly strong effect for a relatively modest (factor of 6.4) change in the disc surface density; the surface densities in real protoplanetary discs are thought to change by factors $\gtrsim 10^3$ over their lifetimes \citep*[e.g.,][]{hartmannetal98,alexanderetal06b}.

The power-law index $\gamma$ also has quite a strong effect on the level of eccentricity growth seen. The two models with $\gamma=1$ ({\sc High25} and {\sc PNMslope}) show slower, weaker growth of eccentricity than their counterparts with the same normalisation value of $\Sigma_0$ but a flat $\gamma=0$ radial power-law dependance.  There are two primary reasons for this. Firstly, a flatter radial profile puts less mass inside the planet's orbit.  The inner disc therefore accretes on to the star more rapidly for flatter surface density profiles (i.e., lower values of $\gamma$), and as eccentricity only begins to grow considerably after the inner disc has accreted, this takes place sooner for flatter surface density profiles. A flatter radial profile also puts more mass into the outer Lindblad resonances, which are responsible for the torques that drive eccentricity growth.  This changes the torque balance on the planet, and gives rise to more rapid eccentricity growth.  In addition, in the special case when Lindblad torques cancel (as proposed by \citealt{goldreichsari03}), the resultant net torque is a function of the surface density gradient, with a steeper power-law dependance damping eccentricity. However, we do not expect this effect to be active in our models, as the planets open gaps sufficiently wide that co-rotation resonances are ineffective.

To investigate this effect further we have followed the method of \citet{artymowiczetal91} to calculate values of $\dot{e}$, time averaged over several orbits at the end of each simulation. The radial contributions to this, normalised against the magnitude of the surface density in each model, are shown in figure \ref{fig9} for all models with $q=0.025$. In the low-surface density limit (models {\sc Low25}, {\sc High25} and {\sc Flat}), we see that the eccentricity is being damped by a peak at $R\sim1.8$, and the magnitude of the surface density acts as a scaling factor with only a very weak dependence on the radial slope ($\gamma$). In these three disc cases, $\dot{e}$ is always negative. By contrast, In the models with higher surface densities, where eccentricity is growing (models {\sc PNM} and {PNMslope}), the sense of the contribution reverses, and the total $\dot{e}$ is positive. Here there is also a more pronounced effect of the radial gradient in the surface density. It is unclear exactly what mechanism causes this effect, but it appears that there is a threshold surface density above which the analysis performed in the low disc mass limit \citep[e.g.][]{goldreichtremaine80} no longer applies.  Our results, and those of previous simulations (see Figure \ref{fig8}), suggest that this threshold is given by
\begin{equation}
\pi \Sigma a^2 > M_{\mathrm p}/C
\label{eq16}
\end{equation}
where $C$ is a constant with a value $C \sim 10$.  Discs with surface densities below this threshold are unable to excite significant eccentricity in the planet's orbit.  This behaviour was suggested by PNM01, and our results support their tentative prediction.  The threshold must presumably depend on several other factors (disc viscosity, $H/R$, etc.), and a complete exploration of this parameter space is beyond the scope of this work.  Nevertheless, our results show that eccentricity is only excited in discs with very high surface densities. 

\section{Discussion}\label{section4}

\subsection{Numerical limitations}\label{section4.1}

The biggest potential numerical problem with this work is that the SPH artificial viscosity may cause spurious eccentricity damping. Artificial viscosity can be especially problematic for shearing-disc type problems such as this, as the differential rotation is often mistaken by the algorithm for a shock.  The effects of artificial viscosity typically scale with the SPH smoothing length $h$, and consequently are a strong function of numerical resolution.  However, we have been able to run our simulations at very high resolution, and given the results of the various tests presented in sections \ref{section2.1.1} and \ref{section3.1.1}, we are confident that the SPH artificial viscosity is not a dominant influence on our results.

We further note that our approximation of a locally isothermal equation of state is somewhat idealised, and in particular may not give an accurate description of the spiral density waves induced in the disc.  \citet{bitschkley10} looked at the evolution of initially eccentric planets in fully radiative discs, and made comparisons to models which used an isothermal approximation. They only found the results to be inconsistent for relatively low planet masses ($\lesssim0.6$ \Mjup). The planet masses considered here are far above this, well into the regime where the isothermal approximation holds, and so we do not expect our use of an isothermal equation of state to introduce significant uncertainties in our results.

Perhaps more of a concern is the viscosity in our simulated discs.  We are confident that the SPH artificial viscosity in our simulations does not dominate our results, but it does set a floor to the range of viscosities we can explore: with current computational capabilities we cannot model discs with $\alpha \lesssim 0.01$\footnote{Note that angular momentum transport by the SPH artificial viscosity scales with the smoothing length $h$ \citep{murray96}, which in 3-D scales as $h \propto N^{1/3}$.  Reducing the numerical viscosity to $\alpha \lesssim 0.001$ would therefore require us to increase the SPH particle number $N$ by a factor $\sim 10^3$, which is not currently feasible.}.  Previous simulations using grid-based methods have adopted a wide range of viscosities and temperature prescriptions, with effective $\alpha$ values ranging from $\alpha \sim 10^{-4}$--$10^{-2}$ \citep[e.g.,][]{papaloizouetal01,dangeloetal06,bitschkley10}.  It is not clear which value(s) of $\alpha$ is most appropriate, but we note that our simulations have a viscosity that is somewhat larger (by a factor $\simeq 2$--3) than the canonical values adopted in previous (mostly 2-D) studies. We note also that values inferred from observations of protoplanetary discs are typically of order $\alpha \sim 0.01$ \citep*[e.g.][]{hartmannetal98,kingetal07}. Presumably the surface density threshold for eccentricity growth must depend on the disc viscosity (and indeed temperature), but unfortunately we are not able to explore this issue further.

It must also be borne in mind that the Navier-Stokes viscosity used in our models is merely a first-order approximation, attempting to mimic the effect globally of a process that occurs far below the scales we are able to resolve -- namely the magnetorotational instability (MRI) thought to drive angular momentum transport in protoplanetary discs \citep{balbushawley91}. By adopting a \citet{shakurasunyaev73} alpha-prescription we implicitly assume that the small-scale effects of turbulence in the disc behave like a viscosity on large scales, but it is not clear whether this approximation holds for length scales $\lesssim H$.  In our simulations the planets open gaps on length scales $\gtrsim H$, so turbulent fluctuations on smaller scales are unlikely to have a strong effect.  However, there is some overlap between the length-scales considered here and the typical scales of MRI turbulence, so we note that our results may not hold if the MRI drives significant power on moderate or large scales (as suggested by recent simulations; \citealt*{simonetal12}).  Detailed investigation of this issue is beyond the scope of this paper, but if the viscous approximation does break down at scales $\sim H$ then it seems likely that eccentricity growth could be affected, particularly for low-mass planets.

\begin{figure}
\includegraphics[width=\columnwidth]{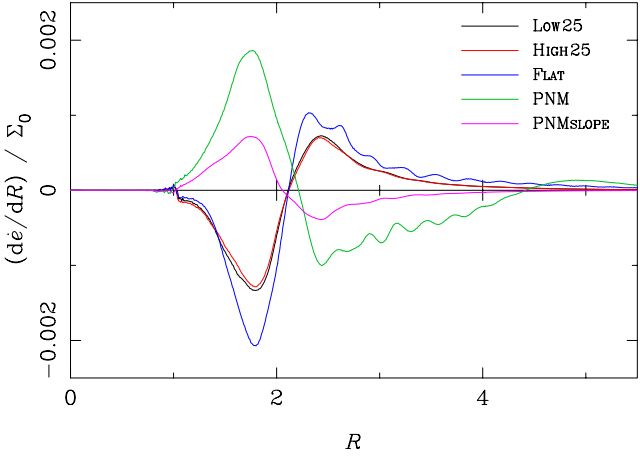}
\caption{Radial contributions to $\dot{e}$, calculated using a Gaussian perturbation approach following the method of \citet{artymowiczetal91}. These values are time-averaged over 5 orbital periods of the planet at the end of each simulation. Models {\sc Low25}, {\sc High25} and {\sc Flat} clearly show that in the low-surface density limit, $\dot{e}$ is approximately linear with the magnitude of the surface density, with a weak dependance on the radial profile $\gamma$. For models {\sc PNM} and {\sc PNMslope} it can be seen that above the threshold for eccentricity growth, the sign of the curves flip and the net $\dot{e}$ becomes positive. In this limit the radial profile of the surface density becomes more important, and the effect is no longer linear with its magnitude.}
\label{fig9}
\end{figure}

\subsection{Applications to real systems}\label{section4.2}

Our major result is that resonant torques are not generally an efficient means of exciting eccentricity, and that planet-disc interactions are unlikely to be responsible for the eccentricities seen in the majority of exoplanet systems.  While we do find eccentricity growth in agreement with \citetalias{papaloizouetal01}, we note that their calculations considered a very massive planet in a disc with a very high surface density.  Their reference surface density of $6.4\times10^4$g/cm$^2$ at 1 au is a factor of several larger than predicted by the Minimum Mass Solar Nebula \citep{weidenschilling77} or more realistic accretion disc models \citep[e.g.,][]{hartmannetal98}. We find that a modest reduction in the disc surface density results in no significant eccentricity growth for similarly massive planets, in agreement with behaviour suggested by \citetalias{papaloizouetal01}. Consequently it is unlikely that this mechanism will be able to excite eccentricity in real protoplanetary discs.

We also fail to find eccentricity growth for lower planet masses, but in this case we treat our results with more caution. Using 2-D simulations \citet{dangeloetal06} found eccentricity growth up to $e\sim0.1$ for planets between 1-3 \Mjup, but typically this occurred on time-scales of thousands of orbits.  Given the high computational cost of our 3-D simulations we are not able to follow their evolution for such long time-scales, and consequently we cannot rule out this slower mode of growth.  However, we note that the disc model used by \citet{dangeloetal06} also uses a very large surface density: if we re-scale their model to match the units used here, their surface density normalisation ($\Sigma_0$) becomes $2.0 \times 10^3$ g/cm$^2$ at 1 au, larger than in our {\sc High} models. We also note that their relatively flat choice surface density profile ($\gamma = 1/2$) is likely to promote eccentricity growth. Although our simulations do not allow us to rule out growth on very long time-scales, analysis like that shown in figure \ref{fig9} shows that the disc gives a negative contribution to the planet's $\dot{e}$ which indicates that eccentricity will not grow unless the disc structure changes significantly. Instead it seems likely that differences in the choice of disc model are responsible for the apparent discrepancy between our results and those of \citet{dangeloetal06}.

As noted in section \ref{section3.2.2}, there are two reasons why the radial surface density profile plays a role in exciting eccentricity growth. At a basic level, a lower value of $\gamma$ puts mass mass into the outer Lindblad resonances and conversely less mass into the inner resonances. As the mode of growth relies on the 1:3 outer resonance \citepalias{papaloizouetal01}, this favours eccentricity growth by strengthening that resonance. In the case of lower mass planets, where a gap is not fully opened, there is another effect which takes place that brings the radial surface density gradient into play. \citet{goldreichsari03} suggest that in a near-Keplerian disc where an outer and inner Lindblad resonance cancel to a reasonable approximation, the resulting net torque scales with $d\Sigma/dr$ rather than with $\Sigma$. This implies that not only the $\Sigma_0$ level but also the radial profile may become as important as the planet mass in discerning the physical contribution of resonant torques for low mass planets.  We believe that it is the former effect that we are seeing in our simulations, with both the surface density profile and its normalisation level both having a strong effect on if, and how, the orbital eccentricity of an embedded planet will grow (figure \ref{fig7}).

This analysis of the effect of both the magnitude and radial profile of the surface density is supported by looking at the radial contributions to the change in eccentricity in different models, shown in figure \ref{fig9}. Below the threshold for eccentricity growth the surface density profile has only a small effect, and its magnitude is an approximately linear scaling factor, which is in agreement with the findings of \citet{artymowiczetal91}. In contrast, above the threshold for growth the sense of the contributions reverse and the difference between models with the same $\Sigma_0$ but different $\gamma$ becomes more pronounced. We are confident that this effect is real, and warrants further study, but detailed investigation is beyond the scope of this paper. Other disc parameters including viscosity and temperature structure must also have an effect on this result, but we have not investigated them here.

We can extend this analysis by taking equation \ref{eq16} to be a necessary condition for eccentricity growth:
\begin{equation}
C\pi\Sigma a^2 \gtrsim M_{\mathrm{p}} 
\label{eq17}
\end{equation}
where $C\sim 10$ is a constant.  However, giant planets continue to accrete via tidal streams even after opening a gap in the disc, and in discs which are sufficiently massive to meet this condition they are likely to accrete rapidly.  This accretion increases $M_{\mathrm p}$, making it less likely that the threshold for eccentricity growth will be met.  The critical issue is therefore one of time-scales: does eccentricity grow on a shorter time-scale than the planet's mass?  For massive giant planets ($\gtrsim 5$\Mjup) tidal torques strongly suppress accretion from the disc on to the planet \citep*[e.g.,][]{lubowetal99}, and we therefore expect eccentricity to grow more rapidly than the planet mass.  However, our results suggest that eccentricity growth in this regime requires very high disc surface densities, $>10^3$g\,cm$^{-2}$ (see Fig.\ref{fig8}), and such massive discs are not commonly observed (see below).

By contrast, lower-mass planets ($\sim 1$\Mjup) accrete very efficiently from their parent discs \citep*[e.g.,][]{dangeloetal02}.  We can parametrize the planetary accretion rate as $\dot{M_{\mathrm p}} = \epsilon \dot{M_{\mathrm d}}$, where $\dot{M_{\mathrm d}} = 3\pi\nu\Sigma$ is the disc accretion rate in the absence of the planet. ÊThe parameter $\epsilon$ represents the efficiency of accretion on to the planet; simulations show that this efficiency has a peak value of $\epsilon \sim 1$ for planets of approximately Jupiter mass, and declines to higher planet masses \citep[][see also \citealt{verasarmitage04}]{lubowetal99,dangeloetal02}. ÊIf we substitute $\nu = \alpha \Omega H^2$ we can rearrange equation \ref{eq16} to find
\begin{equation}
\frac{M_{\mathrm{p}}}{\dot{M_{\mathrm{p}}}} \lesssim \frac{C}{3\epsilon}\alpha^{-1} \Omega^{-1} \left(\frac{H}{a}\right)^{-2}\,.
\label{eq18}
\end{equation}
The quantity on the left-hand-side is the time-scale for planet growth through accretion of gas from the disc, $\tau_{\mathrm{accrete}}$, so we see that meeting the condition for eccentricity growth (equation \ref{eq16}) also sets an upper limit to the time-scale for planet growth by accretion.  Taking standard values of $\alpha = 0.01$ and $H/a = 0.1$, and assuming that $\epsilon \sim 1$, we find
\begin{equation}
\tau_{\mathrm{accrete}} \lesssim 3\times 10^{4}\, \Omega^{-1} \, ,
\label{eq19}
\end{equation}
for planets of approximately Jupiter mass.  This is comparable to the time-scale for eccentricity growth seen in previous studies of Jupiter-mass planets \citep{dangeloetal06}.  We therefore suggest that in this case the planet would accrete rapidly, and `migrate' out of the region of allowed eccentricity growth highlighted in figure \ref{fig8} before it attains a significant eccentricity.

We have shown that in moderately viscous discs (with $\alpha \sim 0.01$), eccentricity growth only occurs if the disc surface density is high.  Unfortunately, observational constraints on the surface densities of discs are extremely weak, and at au scales such as those considered here, almost non-existent. \citet{andrewswilliams07} and \citet{andrewsetal09,andrewsetal10} made systematic studies of protoplanetary discs in Taurus and Ophiuchus, fitting surface density profiles to submillimeter observations of thermal dust emission. However, the angular resolution of such observations means that they can only resolve scales a few tens of au (in the very best cases), and are not sensitive to the region of most interest for planet formation.  If we extrapolate their results down to the unresolved inner disc, their fits yield values of $\Sigma_0$ at au radii between $\sim 30$ and $\sim 1700$ g/cm$^2$, with most values lying at $\sim 700-800$ g/cm$^2$. The radial power-law indices ($\gamma$) they fit to their observations range between $0.4$ and $1.1$, with a clear preference towards the upper end of this range. These correspond approximately to our {\sc High} disc model, and have both lower surface densities and steeper power-law indices than the disc models used by \citetalias{papaloizouetal01} and \citet{dangeloetal06}.  However it must be stressed that extrapolating these sub-mm observations is by no means justified.  It has even been suggested that substantial mass reservoirs may exist in `dead zones' close to the star \citep[e.g.,][]{gammie96,hartmannetal06,zhuetal10}, but there are currently no useful constraints on disc surface densities at au radii.  ALMA may soon provide real constraints on protoplanetary discs with far higher angular resolution, but it will be some time before it is able to probe radii of a few au \citep*[e.g.,][]{cossinsetal10}.  

Despite this, we argue that significant eccentricity growth due to the planet-disc interaction is unlikely given realistic protoplanetary disc conditions. There is an increasing consensus in the literature that this is the case -- while \citet{dangeloetal06} did see eccentricity growth by this method, it did not rise above $\sim0.15$, lower than the observed values of 0.2--0.3 that this effect is invoked to explain. Similarly, the semi-analytic models \citet{moorheadadams08} were unable to reproduce the observed low-eccentricity distribution. Coupled with our findings, it seems that the planet-disc interaction alone is incapable of reproducing the eccentricities seen in exoplanet observations.

This negative result of course begs the question of what the true origin of exoplanet eccentricities is. An emerging consensus seems to be that scattering events are responsible for most if not all of the distribution \citep[e.g.][]{chatterjeeetal08}. This is backed up by a vast number of tightly-packed multi-planet systems observed by Kepler \citep{batalhaetal12} and by a plethora of planetary systems that have clearly undergone strong interactions \citep[e.g. highly inclined and retrograde planets;][]{wrightetal11}.  However, there is some suggestion that the low-eccentricity end of the distribution perhaps cannot be explained by this mechanism \citep{goldreichsari03,jurictremaine08}, and further work in this area is still required.

\section{Conclusions}\label{conclusion}

We have performed high-resolution 3D SPH simulations of giant planets embedded in protoplanetary discs.  For high disc surface densities and planet masses we find that the planet-disc interaction leads to eccentricity growth, in agreement with previous studies.  However, we have shown that for realistic planet masses and disc properties, the planet-disc interaction is incapable of exciting significant orbital eccentricity growth.  While we do not investigate the effect of different disc viscosities, we identify a threshold surface density for eccentricity growth, and note that this threshold is rarely, if ever, met in real systems, except in cases where the timescale for eccentricity growth is comparable to the timescale for mass accretion by the planet.  We conclude from our simulations that in the case of a real giant planet, the interaction with its parent disc is unlikely to yield growth of its orbital eccentricity at measurable levels. Therefore we suggest that the low but non-zero exoplanet eccentricities observed, not accounted for by simulations of planet-planet scattering events, must have some other origin.

\section*{Acknowledgments}
We thank the anonymous referee for useful comments that helped clarify the text. ACD is supported by an Science \& Technology Facilities Council (STFC) PhD studentship.  RDA acknowledges support from STFC through an Advanced Fellowship (ST/G00711X/1). PJA was supported in part by NASA (NNX09AB90G, NNX11AE12G) and by the NSF (0807471).

Theoretical astrophysics research in Leicester is supported by an STFC Rolling Grant.  
This research used the ALICE High Performance Computing Facility at the University of Leicester. Some resources on ALICE form part of the DiRAC Facility jointly funded by STFC and the Large Facilities Capital Fund of BIS.

\bibliography{mnrasmnemonic,references}

\begin{thebibliography}{}

\bibitem[\protect\citeauthoryear{{Alexander}, {Clarke}, \&
  {Pringle}}{{Alexander} et~al.}{2006}]{alexanderetal06b}
{Alexander} R.~D., {Clarke} C.~J.,  {Pringle} J.~E., 2006, MNRAS, 369, 229

\bibitem[\protect\citeauthoryear{{Andrews} \& {Williams}}{{Andrews} \&
  {Williams}}{2007}]{andrewswilliams07}
{Andrews} S.~M.,  {Williams} J.~P., 2007, ApJ, 659, 705

\bibitem[\protect\citeauthoryear{{Andrews} et~al.}{{Andrews}
  et~al.}{2009}]{andrewsetal09}
{Andrews} S.~M., {Wilner} D.~J., {Hughes} A.~M., {Qi} C.,  {Dullemond} C.~P.,
  2009, ApJ, 700, 1502

\bibitem[\protect\citeauthoryear{{Andrews} et~al.}{{Andrews}
  et~al.}{2010}]{andrewsetal10}
{Andrews} S.~M., {Wilner} D.~J., {Hughes} A.~M., {Qi} C.,  {Dullemond} C.~P.,
  2010, ApJ, 723, 1241

\bibitem[\protect\citeauthoryear{{Artymowicz} et~al.}{{Artymowicz}
  et~al.}{1991}]{artymowiczetal91}
{Artymowicz} P., {Clarke} C.~J., {Lubow} S.~H.,  {Pringle} J.~E., 1991, ApJ,
  370, L35

\bibitem[\protect\citeauthoryear{{Balbus} \& {Hawley}}{{Balbus} \&
  {Hawley}}{1991}]{balbushawley91}
{Balbus} S.~A.,  {Hawley} J.~F., 1991, ApJ, 376, 214

\bibitem[\protect\citeauthoryear{{Balsara}}{{Balsara}}{1995}]{balsara95}
{Balsara} D.~S., 1995, J.\ Comput.\ Phys., 121, 357

\bibitem[\protect\citeauthoryear{{Batalha} et~al.}{{Batalha}
  et~al.}{2012}]{batalhaetal12}
{Batalha} N.~M. et~al., 2012, arXiv:1202.5852v1

\bibitem[\protect\citeauthoryear{{Bitsch} \& {Kley}}{{Bitsch} \&
  {Kley}}{2010}]{bitschkley10}
{Bitsch} B.,  {Kley} W., 2010, A \& A, 523, A30

\bibitem[\protect\citeauthoryear{{Chatterjee} et~al.}{{Chatterjee}
  et~al.}{2008}]{chatterjeeetal08}
{Chatterjee} S., {Ford} E.~B., {Matsumura} S.,  {Rasio} F.~A., 2008, ApJ, 686,
  580

\bibitem[\protect\citeauthoryear{{Chiang}, {Fischer}, \& {Thommes}}{{Chiang}
  et~al.}{2002}]{chiangetal02}
{Chiang} E.~I., {Fischer} D.,  {Thommes} E., 2002, ApJ, 564, L105

\bibitem[\protect\citeauthoryear{{Cossins}, {Lodato}, \& {Testi}}{{Cossins}
  et~al.}{2010}]{cossinsetal10}
{Cossins} P., {Lodato} G.,  {Testi} L., 2010, MNRAS, 407, 181

\bibitem[\protect\citeauthoryear{{Cuadra} et~al.}{{Cuadra}
  et~al.}{2009}]{cuadraetal09}
{Cuadra} J., {Armitage} P.~J., {Alexander} R.~D.,  {Begelman} M.~C., 2009,
  MNRAS, 393, 1423

\bibitem[\protect\citeauthoryear{{Cuadra} et~al.}{{Cuadra}
  et~al.}{2006}]{cuadraetal06}
{Cuadra} J., {Nayakshin} S., {Springel} V.,  {Di Matteo} T., 2006, MNRAS, 366,
  358

\bibitem[\protect\citeauthoryear{{D'Angelo}, {Henning}, \& {Kley}}{{D'Angelo}
  et~al.}{2002}]{dangeloetal02}
{D'Angelo} G., {Henning} T.,  {Kley} W., 2002, A\&A, 385, 647

\bibitem[\protect\citeauthoryear{{D'Angelo}, {Lubow}, \& {Bate}}{{D'Angelo}
  et~al.}{2006}]{dangeloetal06}
{D'Angelo} G., {Lubow} S.~H.,  {Bate} M.~R., 2006, ApJ, 652, 1698

\bibitem[\protect\citeauthoryear{{Ford}, {Havlickova}, \& {Rasio}}{{Ford}
  et~al.}{2001}]{fordetal01}
{Ford} E.~B., {Havlickova} M.,  {Rasio} F.~A., 2001, Icarus, 150, 303

\bibitem[\protect\citeauthoryear{{Gammie}}{{Gammie}}{1996}]{gammie96}
{Gammie} C.~F., 1996, ApJ, 457, 355

\bibitem[\protect\citeauthoryear{{Goldreich} \& {Sari}}{{Goldreich} \&
  {Sari}}{2003}]{goldreichsari03}
{Goldreich} P.,  {Sari} R., 2003, ApJ, 585, 1024

\bibitem[\protect\citeauthoryear{{Goldreich} \& {Tremaine}}{{Goldreich} \&
  {Tremaine}}{1980}]{goldreichtremaine80}
{Goldreich} P.,  {Tremaine} S., 1980, ApJ, 241, 425

\bibitem[\protect\citeauthoryear{{Haisch}, {Lada}, \& {Lada}}{{Haisch}
  et~al.}{2001}]{haischetal01}
{Haisch} K.~E., Jr., {Lada} E.~A.,  {Lada} C.~J., 2001, ApJ, 553, L153

\bibitem[\protect\citeauthoryear{{Hartmann} et~al.}{{Hartmann}
  et~al.}{1998}]{hartmannetal98}
{Hartmann} L., {Calvet} N., {Gullbring} E.,  {D'Alessio} P., 1998, ApJ, 495,
  385

\bibitem[\protect\citeauthoryear{{Hartmann} et~al.}{{Hartmann}
  et~al.}{2006}]{hartmannetal06}
{Hartmann} L., {D'Alessio} P., {Calvet} N.,  {Muzerolle} J., 2006, ApJ, 648,
  484

\bibitem[\protect\citeauthoryear{{Juri{\'c}} \& {Tremaine}}{{Juri{\'c}} \&
  {Tremaine}}{2008}]{jurictremaine08}
{Juri{\'c}} M.,  {Tremaine} S., 2008, ApJ, 686, 603

\bibitem[\protect\citeauthoryear{{Kane} et~al.}{{Kane}
  et~al.}{2012}]{kaneetal12}
{Kane} S.~R., {Ciardi} D.~R., {Gelino} D.~M.,  {von Braun} K., 2012, MNRAS,
  425, 757

\bibitem[\protect\citeauthoryear{{Kenyon} \& {Hartmann}}{{Kenyon} \&
  {Hartmann}}{1987}]{kenyonhartmann87}
{Kenyon} S.~J.,  {Hartmann} L., 1987, ApJ, 323, 714

\bibitem[\protect\citeauthoryear{{King}, {Pringle}, \& {Livio}}{{King}
  et~al.}{2007}]{kingetal07}
{King} A.~R., {Pringle} J.~E.,  {Livio} M., 2007, MNRAS, 376, 1740

\bibitem[\protect\citeauthoryear{{Kley} \& {Dirksen}}{{Kley} \&
  {Dirksen}}{2006}]{kleydirksen06}
{Kley} W.,  {Dirksen} G., 2006, A\&A, 447, 369

\bibitem[\protect\citeauthoryear{{Kozai}}{{Kozai}}{1962}]{kozai62}
{Kozai} Y., 1962, AJ, 67, 591

\bibitem[\protect\citeauthoryear{{Lidov}}{{Lidov}}{1962}]{lidov62}
{Lidov} M.~L., 1962, Planet. Space Sci., 9, 719

\bibitem[\protect\citeauthoryear{{Lodato} \& {Price}}{{Lodato} \&
  {Price}}{2010}]{lodatoprice10}
{Lodato} G.,  {Price} D.~J., 2010, MNRAS, 405, 1212

\bibitem[\protect\citeauthoryear{{Lubow}, {Seibert}, \& {Artymowicz}}{{Lubow}
  et~al.}{1999}]{lubowetal99}
{Lubow} S.~H., {Seibert} M.,  {Artymowicz} P., 1999, ApJ, 526, 1001

\bibitem[\protect\citeauthoryear{{Masset} \& {Ogilvie}}{{Masset} \&
  {Ogilvie}}{2004}]{massetogilvie04}
{Masset} F.~S.,  {Ogilvie} G.~I., 2004, ApJ, 615, 1000

\bibitem[\protect\citeauthoryear{{Moorhead} \& {Adams}}{{Moorhead} \&
  {Adams}}{2008}]{moorheadadams08}
{Moorhead} A.~V.,  {Adams} F.~C., 2008, Icarus, 193, 475

\bibitem[\protect\citeauthoryear{{Morris} \& {Monaghan}}{{Morris} \&
  {Monaghan}}{1997}]{morrismonaghan97}
{Morris} J.~P.,  {Monaghan} J.~J., 1997, J.\ Comput.\ Phys., 136, 41

\bibitem[\protect\citeauthoryear{{Murray}}{{Murray}}{1996}]{murray96}
{Murray} J.~R., 1996, MNRAS, 279, 402

\bibitem[\protect\citeauthoryear{{Naoz}, {Farr}, \& {Rasio}}{{Naoz}
  et~al.}{2012}]{naozetal12}
{Naoz} S., {Farr} W.~M.,  {Rasio} F.~A., 2012, ApJ, 754, L36

\bibitem[\protect\citeauthoryear{{Ogilvie} \& {Lubow}}{{Ogilvie} \&
  {Lubow}}{2003}]{ogilvielubow03}
{Ogilvie} G.~I.,  {Lubow} S.~H., 2003, ApJ, 587, 398

\bibitem[\protect\citeauthoryear{{Papaloizou}, {Nelson}, \&
  {Masset}}{{Papaloizou} et~al.}{2001}]{papaloizouetal01}
{Papaloizou} J.~C.~B., {Nelson} R.~P.,  {Masset} F., 2001, A\&A, 366, 263

\bibitem[\protect\citeauthoryear{{Price}}{{Price}}{2004}]{price04}
{Price} D.~J., 2004, Ph.D. thesis, Univ. Cambridge

\bibitem[\protect\citeauthoryear{{Price}}{{Price}}{2007}]{price07}
{Price} D.~J., 2007, Proc.\ Astron.\ Soc.\ Aust., 24, 159

\bibitem[\protect\citeauthoryear{{Price}}{{Price}}{2012}]{price12}
{Price} D.~J., 2012, J.\ Comput.\ Phys., 231, 759

\bibitem[\protect\citeauthoryear{{Pringle}}{{Pringle}}{1981}]{pringle81}
{Pringle} J.~E., 1981, ARA\&A, 19, 137

\bibitem[\protect\citeauthoryear{{Pringle}, {Verbunt}, \& {Wade}}{{Pringle}
  et~al.}{1986}]{pringleetal86}
{Pringle} J.~E., {Verbunt} F.,  {Wade} R.~A., 1986, MNRAS, 221, 169

\bibitem[\protect\citeauthoryear{{Rasio} et~al.}{{Rasio}
  et~al.}{1996}]{rasioetal96}
{Rasio} F.~A., {Tout} C.~A., {Lubow} S.~H.,  {Livio} M., 1996, ApJ, 470, 1187

\bibitem[\protect\citeauthoryear{{Sargent} \& {Beckwith}}{{Sargent} \&
  {Beckwith}}{1987}]{sargentbeckwith87}
{Sargent} A.~I.,  {Beckwith} S., 1987, ApJ, 323, 294

\bibitem[\protect\citeauthoryear{{Shakura} \& {Sunyaev}}{{Shakura} \&
  {Sunyaev}}{1973}]{shakurasunyaev73}
{Shakura} N.~I.,  {Sunyaev} R.~A., 1973, A\&A, 24, 337

\bibitem[\protect\citeauthoryear{{Simon}, {Beckwith}, \& {Armitage}}{{Simon}
  et~al.}{2012}]{simonetal12}
{Simon} J.~B., {Beckwith} K.,  {Armitage} P.~J., 2012, MNRAS, 2808

\bibitem[\protect\citeauthoryear{{Springel}}{{Springel}}{2005}]{springel05}
{Springel} V., 2005, MNRAS, 364, 1105

\bibitem[\protect\citeauthoryear{{Takeda} \& {Rasio}}{{Takeda} \&
  {Rasio}}{2005}]{takedarasio05}
{Takeda} G.,  {Rasio} F.~A., 2005, ApJ, 627, 1001

\bibitem[\protect\citeauthoryear{{Tanaka}, {Takeuchi}, \& {Ward}}{{Tanaka}
  et~al.}{2002}]{tanakaetal02}
{Tanaka} H., {Takeuchi} T.,  {Ward} W.~R., 2002, ApJ, 565, 1257

\bibitem[\protect\citeauthoryear{{Veras} \& {Armitage}}{{Veras} \&
  {Armitage}}{2004}]{verasarmitage04}
{Veras} D.,  {Armitage} P.~J., 2004, MNRAS, 347, 613

\bibitem[\protect\citeauthoryear{{Weidenschilling}}{{Weidenschilling}}{1977}]{weidenschilling77}
{Weidenschilling} S.~J., 1977, Ap\&SS, 51, 153

\bibitem[\protect\citeauthoryear{{Wright} et~al.}{{Wright}
  et~al.}{2011}]{wrightetal11}
{Wright} J.~T. et~al., 2011, PASP, 123, 412

\bibitem[\protect\citeauthoryear{{Wu} \& {Lithwick}}{{Wu} \&
  {Lithwick}}{2011}]{wulithwick11}
{Wu} Y.,  {Lithwick} Y., 2011, ApJ, 735, 109

\bibitem[\protect\citeauthoryear{{Zhu}, {Hartmann}, \& {Gammie}}{{Zhu}
  et~al.}{2010}]{zhuetal10}
{Zhu} Z., {Hartmann} L.,  {Gammie} C., 2010, ApJ, 713, 1143

\end{thebibliography}
\bibliographystyle{mnras}

\label{lastpage}

\end{document}